\documentclass[12pt]{article}

\usepackage{scicite}
\usepackage{graphicx}
\usepackage{amsmath}
\usepackage{url}
\usepackage{subcaption}

\usepackage{times}

\newenvironment{sciabstract}{%
\begin{quote} \bf}
{\end{quote}}

\newcounter{lastnote}

\title{The happiness paradox: your friends are happier than you.}

\author
{Johan Bollen$^{1,3\ast}$ Bruno Gon\c calves$^{2}$, Ingrid van de Leemput$^{1,3}$, and Guangchen Ruan$^{1}$\\
\\
\normalsize{$^{1}$Indiana University, Bloomington IN}\\
\normalsize{$^{2}$Center for Data Science, New York University, New York, NY.}\\ 
\normalsize{$^{3}$Wageningen University, 6700 AA, Wageningen, The Netherlands}\\
\\
\normalsize{$^\ast$To whom correspondence should be addressed; E-mail:  jbollen@indiana.edu}
}

\date{}

\begin{document} 


\baselineskip24pt


\maketitle


\begin{sciabstract}
Most individuals in social networks experience a so-called Friendship Paradox: they are less popular than their friends on average. This effect may explain recent findings that widespread social network media use leads to reduced happiness. However the relation between popularity and happiness is poorly understood. A Friendship paradox does not necessarily imply a Happiness paradox where most individuals are less happy than their friends. Here we report the first direct observation of a significant Happiness Paradox in a large-scale online social network of $39,110$ Twitter users. Our results reveal that popular individuals are indeed happier and that a majority of individuals experience a significant Happiness paradox. The magnitude of the latter effect is shaped by complex interactions between individual popularity, happiness, and the fact that users cluster assortatively by level of happiness. Our results indicate that the topology of online social networks and the distribution of happiness in some populations can cause widespread psycho-social effects that affect the well-being of billions of individuals.
\end{sciabstract}



\paragraph{Introduction}We are a profoundly social species\cite{social:coren2012}. The ability to establish face-to-face, physical relationships in a rich social environment is paramount to our happiness and individual well-being \cite{social:dunbar2010,social:steptoe2013,village:pinker2015}. However, technology is now playing an increasing role in forming our networks of social relationships. Nearly $1/7$th of the world's population and over $2/3$rd of the US population \cite{pew2015:social} now use some form of social media which enables individuals to maintain virtual social networks that extend well beyond geographical, economic, cultural, and linguistic boundaries.

\begin{figure}[h!]
\begin{center}
\includegraphics[width=16cm]{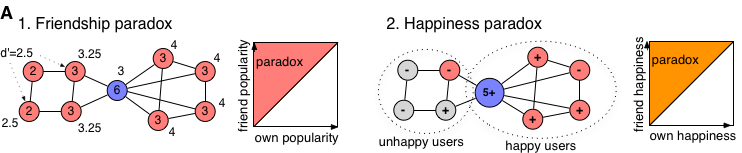}
\caption{\label{schematic} \textbf{A1:} Most social networks are characterized by very skewed degree distributions: a few individuals have very many connections, while most individuals have few connections. The number of connections are marked within each node. Those with many connections are by definition more likely to be someone's friend. As a result their higher number of connections can increase the average degree of given friendship neighborhoods throughout the network (marked above each node) leading to a Friendship paradox (red nodes) in which most individuals nodes are less popular than the average of their friends. \textbf{A2:} When popular individuals are also more likely to be happy, their Happiness becomes more prevalent, raising average happiness levels throughout the friendship circles in the network. A Happiness paradox may result in which most individuals are less happy than their own friends on average. Individuals may cluster based on their Happiness or even the degree to which they experience a Happiness paradox.}
\end{center}
\end{figure}

Evidence has been accumulating that online social networking is associated with elevated levels of loneliness, anxiety, displeasure, and dissatisfaction \cite{Burke:2010:SNA:1753326.1753613,facebook:Kross2013,hpi:facebook2015,geogra:lewis2013}. The reason for this apparent contradiction is unknown, but it may be found in universal social network connectivity patterns. Surprisingly, measured in number of connections, most people will have fewer friends than their own friends do on average \cite{whyyou:feld1991,friends:lerman2013,weirdness:lerman2014}. This phenomenon, commonly referred to as the Friendship Paradox, has been attributed to an inherent structural bias in social network that favors popular individuals: they are by definition more likely to belong to someone's social circle, thereby elevating local levels of popularity. If individuals equate popularity with prestige and compare their own popularity to that of their friends this may lead to increased levels of dissatisfaction (see Fig.\ref{schematic}).

The effects of this Friendship paradox may extend beyond popularity. If popular individuals tend to be happier, their elevated happiness will become more prevalent as well. This may in turn lead to a Happiness paradox\cite{general:jo2014,generalized:eom2014, friends:lerman2013}, where most individuals are less happy than their friends on average (see Fig.\ref{schematic}). In fact, the latter will contribute more directly to the negative psycho-social effects of social networking, since it affects how individuals assess their own Subjective Well-being, i.e. general happiness or life satisfaction \cite{object:oswal2010,progres:krueger2014}, relative to that of others\cite{posel2010:relat,RePEc:cca:wpaper:159}. 
At this point however it has not been established whether (1) popular individuals are indeed happier and (2) a Happiness Paradox does in fact occur in social networks. Given the magnitude of social media adoption these are questions of global importance that may affect the well-being of billions of individuals.

Here we present the first large-scale longitudinal study of happiness and popularity levels for a network of 39,110 Twitter users that are connected by ``friendship'' relations (see Materials and Methods). 
We automatically assess each individuals' Subjective Well-Being (SWB), on a scale of $[-1,+1]$ (See Materials and Methods), by applying a subjective sentiment analysis algorithm to their $3,200$ most recent time-coded Tweets~\cite{happin:Bollen2011}. Their ``Happiness'' (quantified as SWB) along with their ``Popularity'' (quantified as their number of in-network friends) is used to determine: 1) the fraction of individuals that has lower \emph{popularity} than their friends on average ($P_{\text{pop}}$), 2) the fraction of individuals that has lower \emph{happiness} than their friends on average ($P_{\text{hap}}$), and, finally, 3) the correlation between individual happiness and popularity (R(Happiness, Popularity)). 


With these definitions, a Friendship or Happiness Paradox for our sample is indicated by $P_{\text{pop}}$ and $P_{\text{hap}}$ values larger than 50\%, i.e. a majority of individuals have lower Popularity or Happiness than their friends on average. To assess the correlation between Happiness and Popularity we simply calculate Pearson's R correlation between the SWB values and log(degree) of all subjects in our cohort. The use log(degree) is meant to compensate for the very skewed distribution of degree values in our network.

The distribution of subjects and friendship relations in our social network is very unequal, so we assess the robustness of our results by performing a bootstrapping procedure in which we randomly sample 10\% of subjects and their network connections with replacement 5,000 times to assess the distribution of our paradox indicators for different samples of our network. Furthermore, we validate the statistical significance of our results by comparing them to a null-model where we reshuffle the SWB values across all individuals in our network. In this way, we are able to maintain the same identical distribution of SWB values and network structure, while completely eliminating any possible correlation that might be present. The null-model was bootstrapped 20,000 times and, as expected, it eliminated the Happiness paradox.

\begin{table}
\begin{center}
\begin{tabular}{c||l}
					    &	Value (\%) and 95\% CI				       \\\hline
$P_{\text{pop}}$			&	94.3\%, [93.7, 94.9]	N=5,000			\\
$F_{happ}	$			    &	58.5\%, [57.2, 59.8]	N=5,000			\\
Null-model			        &	50.1\%, [49.4, 50.8]    N=20,000		\\\hline
R(Happiness, Popularity)	&	0.109, [0.076, 0.140]	N=5,000			\\\hline
\end{tabular}
\caption{\label{paradox} Magnitude of Friendship Paradox, Happiness Paradox (compared to null-model produced by randomly re-assigning SBW values across all subjects), and Happiness-Popularity correlation coefficient (Pearson's R) for all subjects (N=39,110).}
\end{center}
\end{table}

As shown in Table \ref{paradox}, we find that $P_{\text{pop}}=94\%$ indicating a very significant Friendship Paradox across all subjects, meaning that the great majority of users are less popular than their friends are on average. We also find a modest but robust value of $P_{\text{hap}}=59\%$, indicating the presence of a Happiness Paradox. Hence a majority of subjects is indeed less happy than their friends on average. Our null-model indicates the absence of a Happiness paradox when the effects of network structure on Happiness levels are removed by random re-assignment. The lower magnitude of the Happiness Paradox could result from the rather low yet robust correlation between Happiness and Popularity (Spearman's $R=0.100$).

\begin{figure}[h!]
\begin{center}
\includegraphics[width=16cm]{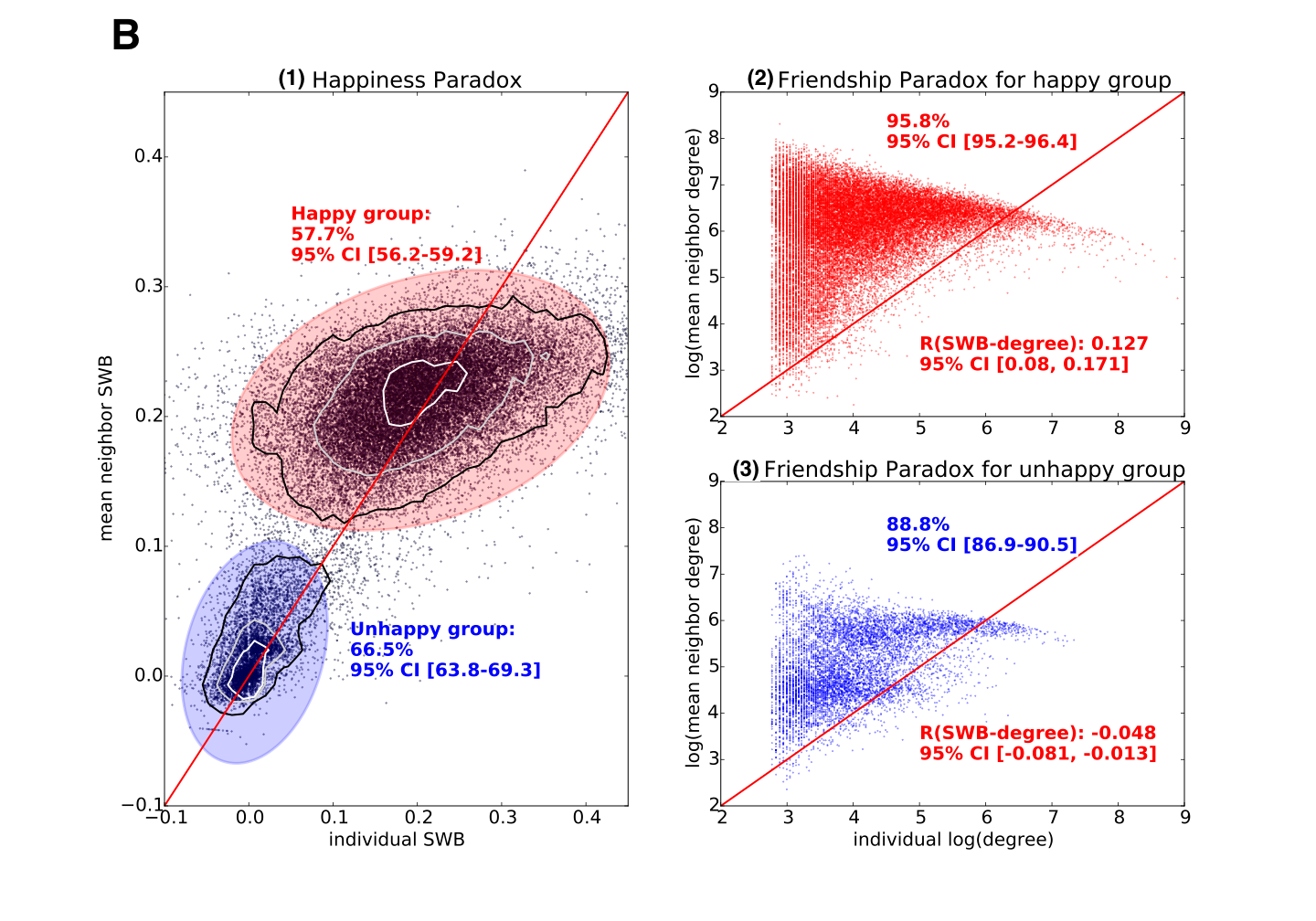}
\caption{\label{diagonal_contour} \textbf{(B1) Happiness Paradox}: Distribution of individual Happiness (x-axis) vs. average Happiness of one's friend's average (y-axis). Happiness is measured in terms of longitudinal Subjective Well-Being (SWB) scores. Subjects above the red paradox line experience lower happiness (SWB) than their friends' average. The distribution of SWB scores places a majority of subjects well above the diagonal Paradox line. Ellipses indicate the boundaries of 2 Gaussian Mixture Model components used to demarcate a Happy (red) and Unhappy (blue) groups of subjects. Paradox magnitudes are expressed in terms of the percentage of users who experience lower happiness than their friends. The 95\% confidence intervals are calculated by a 5000-fold bootstrapping of a 10\% sample to determine the sensitivity of our results to random network sampling variations. \textbf{(B2) and (B3) Friendship Paradox}: Distribution of individual Popularity (x-axis) vs.~average Popularity of one's Friends (y-axis). Popularity is measured in terms of Log(degree) in the Friendship network. Subjects above the red paradox line experience lower popularity than their friends on average.
As shown, we find significant Happiness and Friendship Paradoxes for all users, but Happy users experience a stronger Friendship Paradox whereas Unhappy users experience a stronger Happiness Paradox.}
\end{center}
\end{figure}
The distribution of individual Happiness levels and mean neighborhood happiness in our sample is distinctly bi-modal. This is congruent with the observed distribution of Subjective Well-being across several cultures and nations \cite{measure:balaton2011}. As shown in Fig. \ref{diagonal_contour} this bi-modality also occurs at the level of our friendship network which separates subjects into 2 distinct groups: happy subjects with happy friends (the ``Happy'' group) and unhappy subjects with unhappy friends (the ``Unhappy'' group). This result follows earlier reports of happiness being homophilic or assortative in social networks \cite{happin:Bollen2011,McPherson2001,fowler:dynamic2008}. 
Since a Happiness Paradox specifically compares individual happiness to the average happiness of one's friends, this homophilic bi-modality must be factored into our analysis. By performing a separate analysis for Happy and Unhappy groups of users, we attempt to equalize the effects of neighbor happiness across the two groups.

As shown in Fig. \ref{diagonal_contour} we use a Gaussian Mixture Model (GMM) to demarcate our Happy and Unhappy groups. We determine the location and distribution of two separate Gaussian components in the distribution of individual happiness vs.~mean friend happiness (See Suppl. Mat.) and demarcate both groups by simply determining whether a subject and its neighbors fall within 2 standard deviations from the center of one of the components (indicated by 2 ellipses in Fig. \ref{diagonal_contour}). Note that this procedure assumes a Gaussian density distribution which roughly matches the quantiles of the empirical density as shown by the contour lines of Fig. \ref{diagonal_contour}.

%
%

\begin{figure}[h!]
\begin{center}
\includegraphics[width=0.8\linewidth]{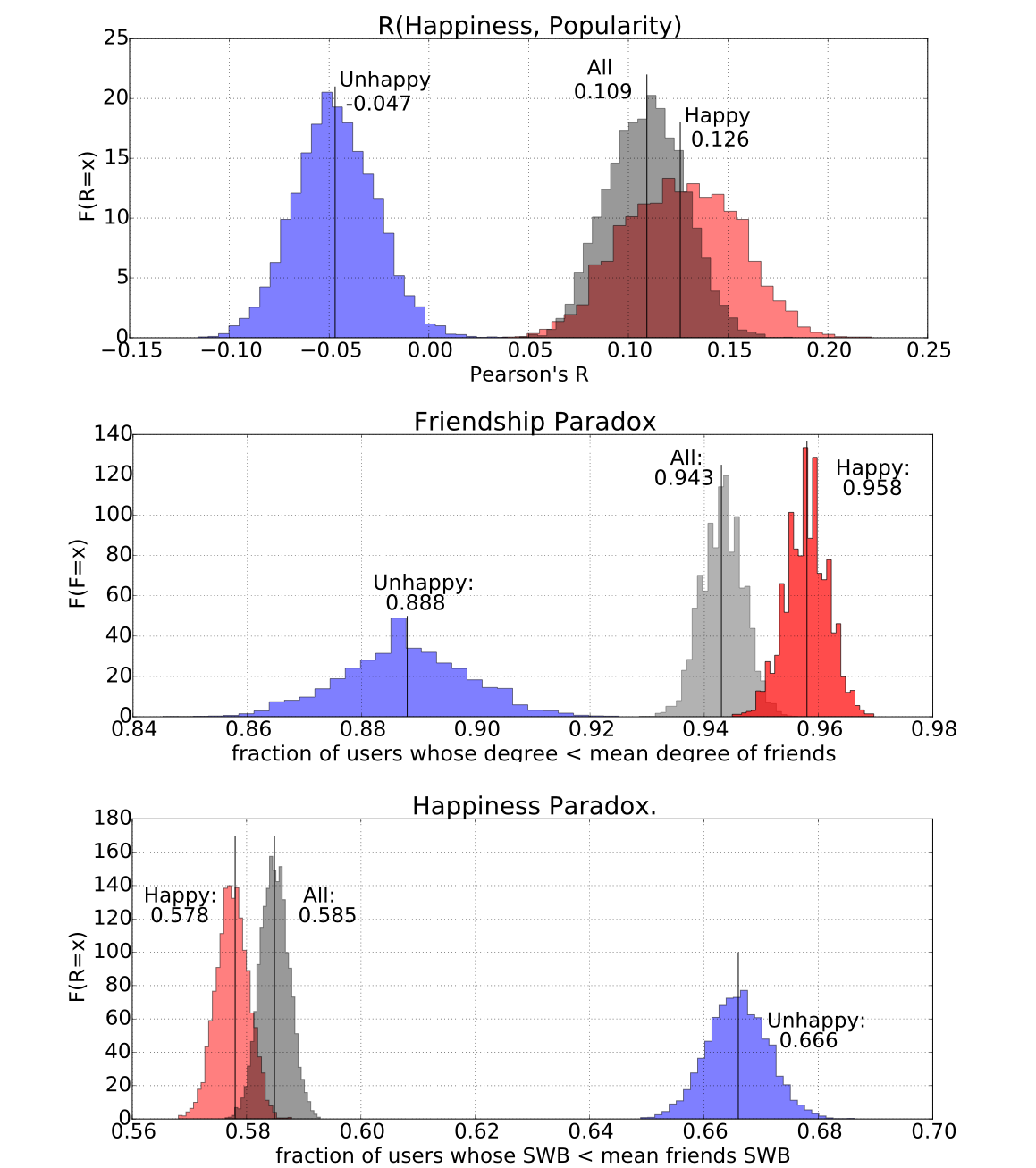}
\caption{\label{all_histograms} \textbf{Top:} Estimated Pearson's R correlation coefficients (95\% Confidence Intervals in brackets) between individual Happiness (Subjective Well-Being) vs. individual Popularity (log degree) for All subjects: 0.109 [0.077, 0.140], Happy group: 0.126 [0.081, 0.171], and unhappy group: -0.047 [-0.08, -0.013]
. \textbf{Middle:} Distribution of Friendship Paradox values for all subjects 0.943 [0.937, 0.949],
happy group: 0.958 [0.951, 0.964], and unhappy group 0.888 [0.869, 0.906]. \textbf{Bottom:} Distribution of Happiness Paradox values for all subjects: 0.585 [0.581, 0.589], happy group: 0.578 [0.573, 0.582], and unhappy group 0.666 [0.657, 0.674]. }
\end{center}
\end{figure}

We re-run our analysis for the Happy and Unhappy groups separately. The results are summarized in Figs. \ref{diagonal_contour} and \ref{all_histograms}. 

	
	
	
	

These results reveal that the Happy group experiences a strong Friendship Paradox but a weak, yet very robust Happiness Paradox. The Unhappy group experiences a weaker Friendship Paradox, but a significantly stronger Happiness Paradox than the Happy group, in spite of subjects being surrounded by less Happy friends. 

To determine whether the strong Happiness Paradox for the unhappy group, in spite of its lower correlation between Popularity and Happiness, may be related to interpersonal effects, such as contagion or increased homophily, we examine the relation between individual happiness and the average happiness of one's neighbors. As indicated  by the distribution of individuals in Fig. \ref{diagonal_contour}, the strength of the relationship between a subjects' Happiness and the average Happiness of their friends differs significantly between the Happy and Unhappy group. A linear regression analysis indicates that individual Happiness and average friends' Happiness are more strongly related within the Unhappy group (b=+0.9439, F=2.222e+05, p$<$0.001) than within the Happy group (b=+0.4459, F=34665, p$<$0.001). This result suggests that unhappy users are more strongly affected by the lower happiness of their friends, possibly explaining why this group exhibits a stronger Happiness Paradox in the absence of a strong correlation between Popularity and Happiness.

\section*{Conclusion}

This work constitutes the first direct measurement of a Happiness paradox in social networks, rather than its theoretical derivation from hypothetical network attributes and properties. Our results suggest that previous observations of decreased happiness among social media users may result directly from a widespread inflated perception of the happiness of one's friends. Although happy and unhappy groups of subjects are both affected by a significant happiness paradox, unhappy subjects are most strongly affected. This is counter-intuitive for two reasons. First, the correlation between happiness and popularity is lowest for individuals in the unhappy group. A happiness paradox can result from a friendship paradox when popularity and happiness are correlated, since more popular and thus more prevalent individuals will increase the average happiness of one's circle of friends. As a result, the unhappy group, with the lowest correlation between popularity and happiness, should experience the lowest happiness paradox. Second, the strong assortativity of happiness in our social network reduces the prevalence of happy subjects in the social network circle of unhappy subjects. Therefore, it should be easier for individuals in this group to surpass the average happiness of their friends. Our results show that neither is the case. A possible explanation may lie in the stronger relation between the happiness of individuals in this group and the overall happiness of their friends. This effect may point to an alternate origin for the occurrence of a Happiness paradox; instead of resulting from the greater prevalence of popular and happy individuals, in some cases, a happiness paradox may result from the complex social interactions between individuals and their friends, e.g.~through mood contagion \cite{experi:kramer2014,ferrara:measur2015,ferrara:quant2015} and potentially verbal commiseration and mirroring.

Our study has limitations. First, the assessment of Subjective Well-Being from social media using text analysis algorithms may not be perfectly reliable. However, given the large number of individuals in our dataset, no indication of consistent directional bias, and the magnitudes of the observed effects, we expect this will not affect the validity of our observations. Future improvements in sentiment and mood analysis, and ground truth obtained from user surveys, may increase the reliability of our SWB estimates. Second, given the large role that social media plays in the social lives of billions of individuals, we expect that these environments may induce longitudinal changes in the public's social behavior and may over time alter the very nature of social relations themselves \cite{social:dunbar2012}. Further analysis will be required to determine the extent and significance of these changes, and how they affect the propensity of online users to experience the effects of a Friendship and Happiness Paradox over time.

In spite of these limitations our results provide a strong indication that widespread social media use may lead to increased levels of social dissatisfaction and unhappiness since individuals will be prone to unfavorably compare their own happiness and popularity to that of others. Happy social media users will likely think their friends are much more popular and slightly happier than they are while unhappy social media users will likely have unhappy friends that will still seem much happier and more popular than they are on average.
We caution against the widespread use of social media given the likelihood that it decreases the happiness and well-being of particularly the most vulnerable groups in society.

\section*{Acknowledgements}
BG thanks the Moore and Sloan Foundations for support as part of the Moore-Sloan Data Science Environment at NYU.

\bibliography{hparadox}
\bibliographystyle{Science}







\clearpage


\section*{Supplementary materials}



\subsection*{The Friendship Network}
To generate a friendship network among Twitter users we started with an initial set of $104,115$ users, for which we downloaded the full list of users that they ``follow'' or that they are ``followed'' by. Reciprocal ``Follow'' and ``Following'' ties are taken as an indication of a friendship relation between the two individuals~\cite{goncalves11-2}. This resulted in a network of $104,115$ node connected by $23,640,257$ reciprocal edges. We further remove subjects with less than $15$ friends in order to improve the reliability of calculating the mean degree and mean SWB values of an individuals' friends. This reduces our final cohort to $39,110$ subjects connected by reciprocal friendship relation.


\subsection*{Friendship Paradox}
 We then assess the magnitude of the Friendship Paradox in our network by calculating the fraction of users $|| u_i \in U||$ whose Popularity, denoted $D\left(u_i\right)$ is lower than the average Popularity of their nearest neighbors (or ``friends'') $NN\left(u_i\right) \in U$, denoted $\mathcal{D}$, vs.~the total number of individuals in the network $||U||$. This yields the magnitude of the Friendship Paradox as:

\begin{equation}
\label{PD}
P\left(D\right) = \frac{|| \{u_i \in U: D\left(u_i\right) < \mathcal{D}\left(NN\left(u_1\right)\right) \} ||} {||U||}
\end{equation}

When the magnitude of $P\left(D\right) > 0.5$ we conclude that the majority of users experiences a Friendship. 

\subsection*{Happiness Paradox}
For each of the $N=39,110$ users that fulfill all the requirements listed above, we further collect their complete Twitter history based on which we can assess the SWB of each individual. With this information hand, the magnitude of the Happiness Paradox can be obtained in a way similar to the way in which we measure the Friendship Paradox. We simply calculate the fraction of users $|| u_i \in U||$ whose Happiness, denoted $H\left(u_i\right)$ respectively, is lower than the average Happiness of their nearest neighbours, denoted $\mathcal{H}$, vs.~the total number of individuals in the network $||U||$, or, mathematically:

\begin{equation}
\label{PH}
P\left(H\right) = \frac{|| \{u_i \in U: H\left(u_i\right) < \mathcal{H}\left(NN\left(u_1\right)\right) \} ||} {||U||}
\end{equation}

\subsection*{Bootstrapping}

To determine the significance of our results we employ a bootstrapping procedure in which we repeatedly re-sample the set of individuals in our data with replacement and re-calculate our indicators to assess the variance of results resulting from random changes in the underlying population. This procedure allows us to obtain Confidence Intervals for all indicators by determining the 5th and 95th percentile of the results obtained for each of 5000 sub-samples with replacement over the entire set of individuals.

\subsection*{Null model}
As mentioned in the text we verify the importance of popularity-happiness correlations by comparing the results we obtained in our dataset with those of a simple null-model. We keep the structure of the network and SWB distributions intact by simply resampling the complete set of SWB values with replacement and re-calculating Eq.~\ref{PH} and Eq.\ref{PD}. This procedure is performed 20,000 times. We report the 95\% confidence intervals for the resulting distribution of paradox values. 

\subsection*{Gaussian Mixture Components}
Our data contains two data points for each user:
\begin{itemize}
    \item their own Popularity or Happiness
    \item the average Popularity or Happiness of their friends
\end{itemize}

Each user can then be described as a point on a 2-dimensional euclidean plane $P$ spanned by their own popularity or happiness (x) and the average happiness of their group of friends (y).

In this plane, users cleanly separate in 2 clusters in $P$ according to matching levels of popularity or happiness. To determine an objective demarcation criterion we use a Gaussian Mixture Model (GMM) to identify membership in either of the 2 groups. The GMM is trained from our empirical data by means of a standard Expectation-Maximization procedure to identify two 2D Gaussian distributions that are each characterized by a center $\mu_c$ and co-variance $\sigma_c$ to best match the distribution of individuals in $P$. Each components carries a weight $w$ with which to mix the 2 components to match the probability density function of the data, but we are only concerned with their location to demarcate the two groups of individuals. The gaussian parameter values obtained using the Scikit-learns sklearn.mixture package without any constraints on the covariance model are:w

\begin{center}
\begin{tabular}{c||c|c}
Component           &   $\mu_c$         &     $\sigma_c$ \\\hline\hline 
1 (Happy group)     &   (0.2037652, 0.21266452)  &       (0.00186789, 0.00046923) \\
2 (Unhappy group)   &  (0.00704093,0.0182976)    &       (0.00046923, 0.0018294)\\\hline
\end{tabular}
\end{center}

\end{document}